\documentclass[3p,times,procedia]{elsarticle}
\usepackage{nupha_ecrc}
\usepackage[hidelinks]{hyperref}
\usepackage{overpic}
\usepackage{lineno}
\usepackage[none]{hyphenat}
\usepackage{color}
\usepackage{capt-of}
\usepackage{array}
\usepackage{amsmath}
\volume{00}

\firstpage{1}

\journalname{Nuclear Physics A}

\runauth{Jasper Parkkila for the ALICE Collaboration}

\jid{nupha}

\jnltitlelogo{Nuclear Physics A}

\usepackage{amssymb}
\usepackage[figuresright]{rotating}

\begin{document}

\begin{frontmatter}


%\title{Linear and non-linear flow modes of charged and identified particles in Pb--Pb collisions at $\sqrt{s_\mathrm{NN}}$=5.02 TeV with ALICE\tnoteref{label1}}
%\tnotetext[label1]{}
%\author{Jasper Parkkila for the ALICE Collaboration\corref{cor1}\fnref{label2}}
\ead{jasper.parkkila@cern.ch}
%\fntext[label2]{University of Jyväskylä}

%% Instructions from Editor: Please use the following \dochead only in the preprint version (e-print arXiv etc.); 
%% use empty \dochead{} when submitting to Nuclear Physics A!
\dochead{XXVIIIth International Conference on Ultrarelativistic Nucleus-Nucleus Collisions\\ (Quark Matter 2019)}
%% e.g. \dochead{17th International Conference on Dynamical Processes in Excited States of Solids}

\title{Linear and non-linear flow modes of charged and identified particles in Pb--Pb collisions at $\sqrt{s_\mathrm{NN}}$=5.02 TeV with ALICE}


\author{Jasper Parkkila for the ALICE Collaboration\corref{cor1}}

\address{CERN, Route de Meyrin, 1211 Geneva, Switzerland}

\begin{abstract}
The higher order harmonic flow observables $v_n$ ($n > 3$) and their non-linear responses to the initial state anisotropy have the strong potential to constrain shear and bulk viscosity to entropy ratios because of different sensitivities for various stages of heavy-ion collisions. The measurements of the flow coefficients and the non-linear coefficients up to the ninth and fifth harmonic, respectively, are presented in Pb--Pb collisions at $\sqrt{s_\mathrm{NN}}$=5.02 TeV for charged hadrons. In addition, the results of $p_\mathrm{T}$-differential non-linear flow modes for $\pi^\pm$, $\mathrm{K}^\pm$, $\mathrm{p}+\mathrm{\bar{p}}$, $\mathrm{K}_s^0$, $\Lambda+\bar{\Lambda}_s^0$ and $\phi$ are presented. The results are compared to the same measurements at 2.76 TeV and calculations from state of the art hydrodynamic models.
\end{abstract}


\end{frontmatter}

\renewcommand{\floatpagefraction}{0.99}%

% \linenumbers

\section{Introduction}
\label{sec:intro}

%\lipsum[1-2]
%The study of ultrarelativistic heavy-ion collisions
%The state of the matter 
\noindent Quark-Gluon Plasma (QGP), the state of the matter that exists at extremely high temperatures and energy densities, is studied at the LHC~\cite{Voloshin:2008dg}. %In an heavy-ion event, a hydrodynamic expansion of the matter is observed. This expansion is driven by the large pressure gradients in the plasma, and can be identified as anisotropic in nature. 
%The anisotropy of the expansion, usually referred to as anisotropic flow, is known to originate from the initial state spatial anisotropies in the energy density distribution.
%As a result, anisotropic flow measurements through multi-particle azimuthal correlations can probe initial state fluctuations and important properties of the shear viscosity to entropy density ratio ($\eta/s$), bulk viscosity to entropy ratio ($\zeta/s$) and equation of state.
One of the primary goals of heavy-ion collision programs is to study the properties of this strongly interacting matter by measuring observables, such as the anisotropic flow that arises from the hydrodynamic expansion of the initial state spatial anisotropies in the energy density profile.
As a result, anisotropic flow measurements through multi-particle azimuthal correlations can probe initial state fluctuations and important properties of the shear viscosity to entropy density ratio ($\eta/s$), bulk viscosity to entropy ratio ($\zeta/s$) and equation of state.

The second and third harmonic coefficients of the anisotropic flow are known to follow an approximately linear relation to their corresponding eccentricity in the initial conditions, with $v_n\propto\varepsilon_n\,(n=2,3)$, where $v_n$ is the magnitude of the $n^\mathrm{th}$ harmonic flow vector $V_n$. However, in higher harmonics $n>3$ this relation no longer holds as the higher harmonics $V_n$ are induced not only by their corresponding $n$-th order eccentricity vector, but also lower order harmonics. It has been shown~\cite{Gardim:2011xv} that the higher order flow can be expressed as a combination of linearly correlated contribution $V_{n\mathrm{L}}$ and one or more contributions from the lower order harmonics:
\begin{equation}
\label{eq:nldecomp}
\begin{split}
	V_4&=V_{4\mathrm{L}}+\chi_{4,22} V_2^2,\\
	V_5&=V_{5\mathrm{L}}+\chi_{5,23} V_2 V_3,\\
	V_6&=V_{6\mathrm{L}}+\chi_{6,222} V_2^3+\chi_{6,33} V_3^2+\chi_{6,24} V_2 V_{4L},\\
	%V_7&=V_{7\mathrm{L}}+\chi_{7,223} V_2^2 V_3+\chi_{7,34} V_3 V_{4L}+\chi_{7,25} V_2 V_{5L}.\\
	\dots
\end{split}
\end{equation}
where $\chi_{n,mk}$, called non-linear flow mode coefficient, characterizes the non-linear flow mode induced by the lower order harmonics. The magnitude of non-linear mode for the fourth harmonic, $v_{4,22}\equiv\chi_{4,22}\sqrt{\langle v_2^4\rangle}=\frac{\Re\langle V_4 (V_2^*)^2\rangle}{\langle v_2^4\rangle^\frac{1}{2}}$ is obtained with a projection of $V_4$ onto the second harmonic plane $\Psi_2$. %expression for $\chi_{n,mk}=\frac{\Re\langle V_4 (V_2^*)^2\rangle}{\langle v_2^4\rangle}$ can be obtained with a projection of Eq.~\ref{eq:nldecomp} onto the lower harmonics. %The resulting expression for the fourth harmonic $\chi_{4,22}$ is then
%\begin{equation}
%\label{eq:chi}
%	%\chi_{4,22}=\frac{v_{4,22}}{\sqrt{\langle v_2^4\rangle}}.
%	\chi_{4,22}=\frac{\Re\langle V_4 (V_2^*)^2\rangle}{\langle v_2^4\rangle}.%\approx\langle v_4\cos(4\psi_4-4\psi_2)\rangle,
%\end{equation}
The measurement of this quantity is performed using the subevent method, where the event is divided into two subevents separated by a pseudorapidity gap $|\Delta\eta|>0.8$. The subevent method effectively suppresses the non-flow contributions from short-range correlations unrelated to the common symmetry plane. For subevent A, the multi-particle correlation is $v_{4,22}^{\mathrm{A}}=\langle\langle\cos(4\varphi_1^\mathrm{A}-2\varphi_2^\mathrm{B}-2\varphi_3^\mathrm{B})\rangle\rangle\allowbreak{}/\langle\langle\cos(2\varphi_1^\mathrm{A}+2\varphi_2^\mathrm{A}-2\varphi_3^\mathrm{B}-2\varphi_4^\mathrm{B}\rangle\rangle^\frac{1}{2}$, while for subevent B the '$\mathrm{A}$' and '$\mathrm{B}$' in the aforementioned expression are interchanged. The final result is the average of the results from the two subevents. In a $p_\mathrm{T}$-differential analysis, $\varphi_1$ is taken from a certain $p_\mathrm{T}$ region ($\varphi_1(p_\mathrm{T})$) or is of a specific particle species.

In these proceedings the measurements of the higher order flow up to $v_9$ are reported in Pb--Pb collisions at $\sqrt{s_\mathrm{NN}}=5.02\,\mathrm{TeV}$~\cite{Acharya:2020taj}. Furthermore, measurements of the non-linear flow mode coefficients $\chi_{n,mk}$ are presented up to the fifth harmonic, among with comparisons to various state of the art hydrodynamical calculations. The high-order harmonic coefficients are expected to have good sensitivities to the hydrodynamic parameters such as $\eta/s$ and $\zeta/s$, and thus improve the constraints on these transport properties. The $p_\mathrm{T}$-differential non-linear flow modes are presented for the identified $\pi^\pm$, $\mathrm{K}^\pm$, $\mathrm{p}+\mathrm{\bar{p}}$, $\mathrm{K}_s^0$, $\Lambda+\bar{\Lambda}_s^0$ and $\phi$~\cite{Acharya:2019uia}. Such measurements yield additional constraints for the initial conditions, $\eta/s$ and $\zeta/s$, and revealing information about different particle production mechanisms.

%
%The initial state parameterizated through a cumulant defined eccentricity 

%The relation between the ... is well known to be approximately linear 

\section{Analysis Details}
\label{sec:analysis}

\noindent The data sample consists of about 42 million minimum-bias Pb--Pb collisions at $\sqrt{s_\mathrm{NN}}=5.02\,\mathrm{TeV}$, recorded by ALICE~\cite{Aamodt:2008zz} in 2015. The trigger requires coincidence of signals from the two scintillator arrays, V0A and V0C~\cite{Aamodt:2008zz}. The track reconstruction is based on information from the Time Projection Chamber (TPC)~\cite{Aamodt:2008zz} and the Inner Tracking System (ITS)~\cite{Aamodt:2008zz}. For unidentified flow, only particle tracks within the transverse momentum interval $0.2<p_T<5.0\,\mathrm{GeV}/c$ and pseudorapidity range $0.4<|\eta|<0.8$ are considered. To suppress the non-flow, a pseudorapidity gap of $|\Delta\eta|>0.8$ is used. For the analysis of identified hadrons, the TPC and Time-Of-Flight (TOF) detectors are used to identify the pions ($\pi^\pm$), kaons ($\mathrm{K}^\pm$), (anti-)protons ($\mathrm{p}+\mathrm{\bar{p}}$). For decaying particles $\mathrm{K}_s^0$, $\Lambda+\bar{\Lambda}_s^0$ and $\phi$, the invariant mass method is employed. A minimum 80\% purity is maintained for all particle species. The sub-event method (without $\eta$-separation) is applied and the possible remaining non-flow is considered in the systematical uncertainty. The observables in this analysis are measured with multi-particle correlations obtained using the generic framework~\cite{Bilandzic:2013kga} for anisotropic flow analysis.

\section{Results}
\label{sec:results}

\begin{figure}[tbp]
	\begin{minipage}{0.5\textwidth}
	\centering
	\hspace{-2em}\begin{overpic}[width=0.87\textwidth]{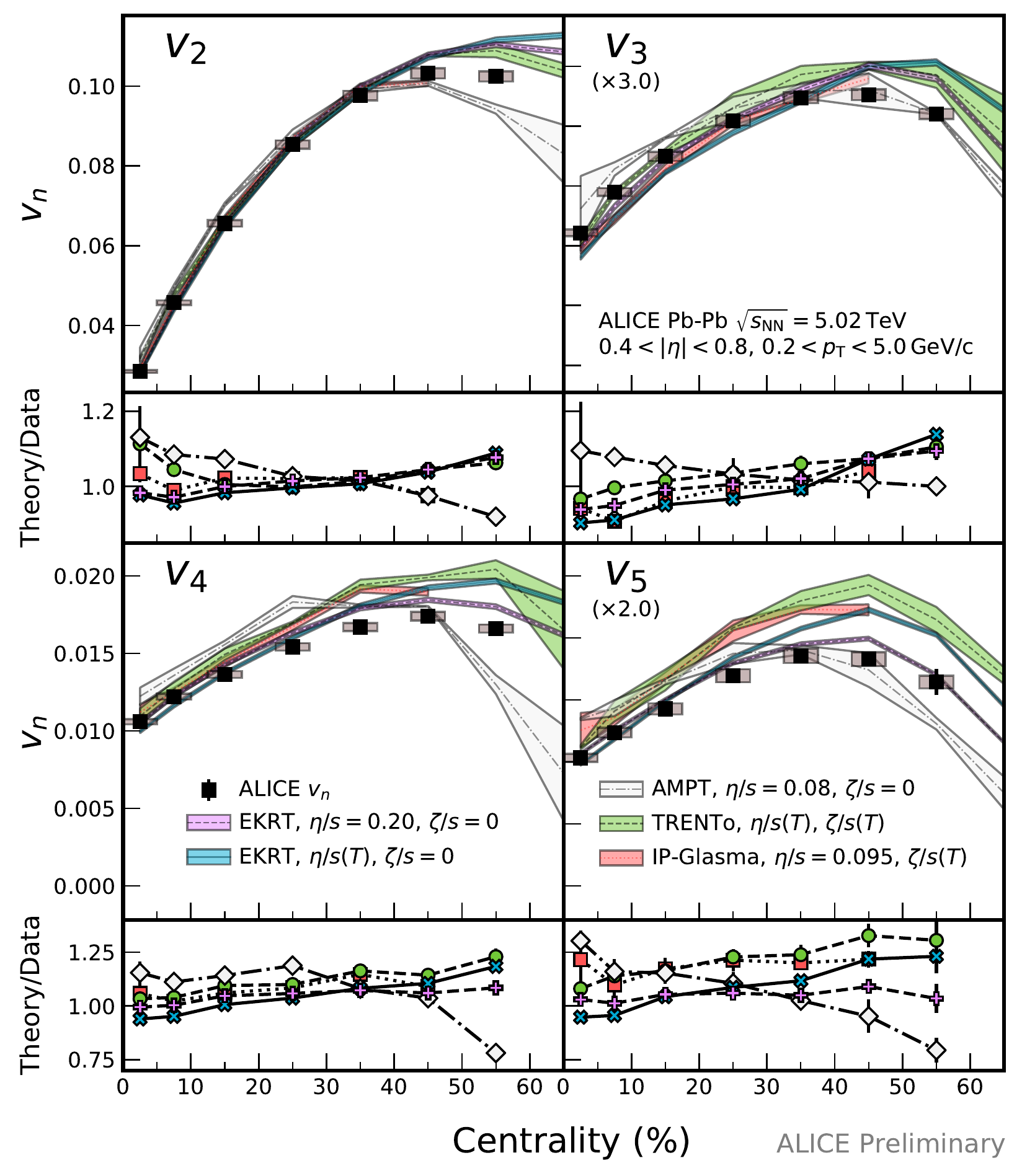}
		\put(0,1.5){\includegraphics[width=0.3\textwidth]{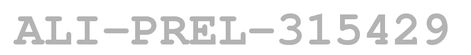}}
	\end{overpic}
	\caption{$v_n$ with hydrodynamical model calculations.}
	\label{fig:vn}
	\end{minipage}
	\begin{minipage}{0.5\textwidth}
	\centering
	\hspace{-3em}\begin{overpic}[width=0.87\textwidth]{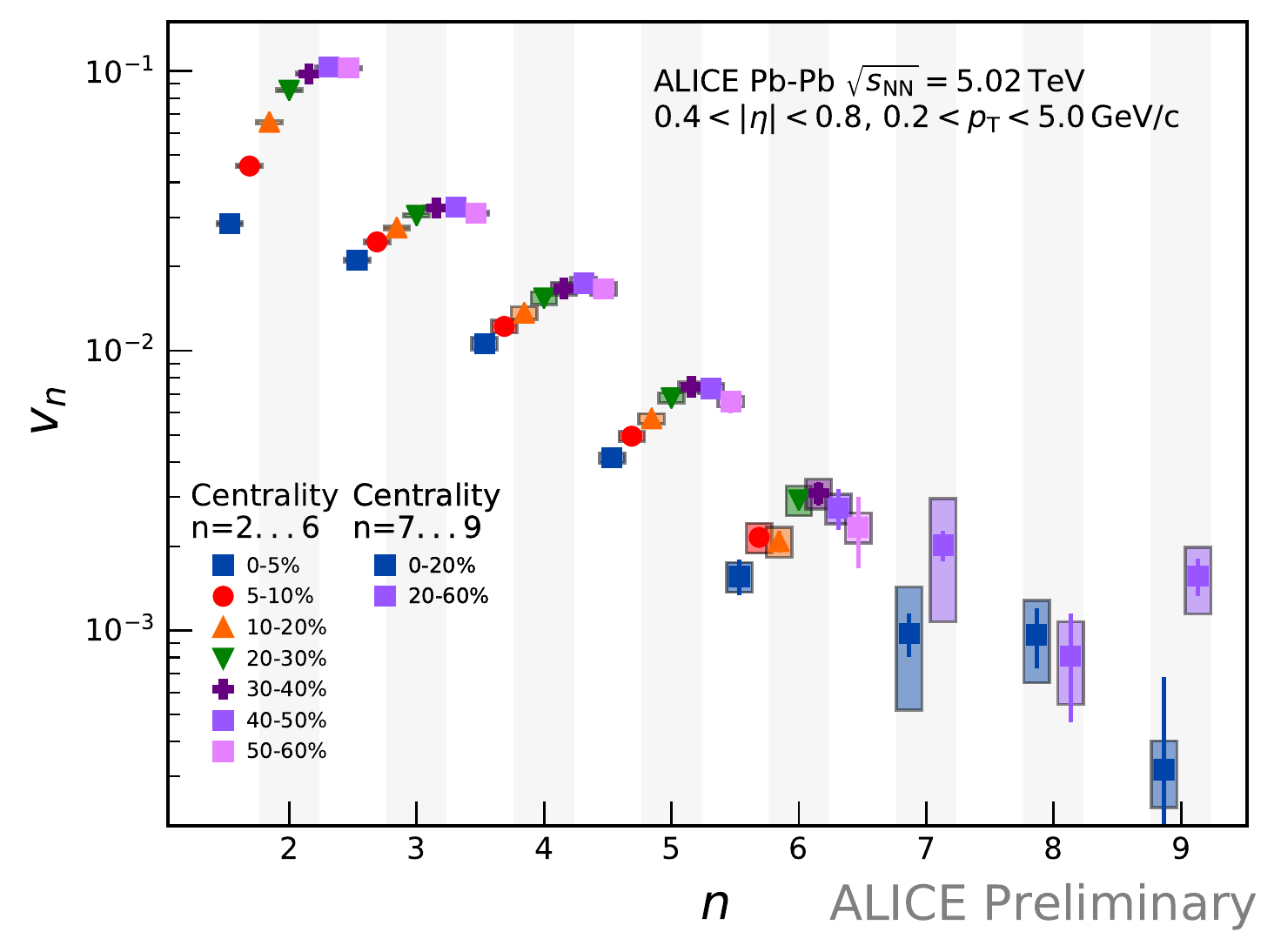}
		\put(7,1){\includegraphics[width=0.3\textwidth]{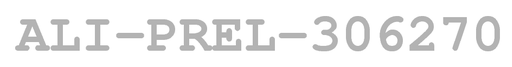}}
	\end{overpic}
	\caption{$v_n$ as a function of the harmonic order $n$ for various centrality intervals.}
 	\label{fig:pspectra}
	\end{minipage}%
\end{figure}

\newcommand{\trento}{T\raisebox{-.5ex}{R}ENTo}

\noindent In Fig.\ref{fig:vn} the flow coefficients are presented up to $v_5$ among with hydrodynamical calculations. IP-Glasma with viscous hydrodynamics describes the data best for $n=2$ and $n=3$, while EKRT($\eta/s=0.2$) has the best agreement for $n=4$ and $n=5$. Likewise, \trento{} has a good agreement for $n=2$ and $n=3$, but overestimates the higher harmonics. AMPT predicts the data better in peripheral collisions, but fails to describe the data in central collisions for higher harmonics. %As the harmonic increases, the sensitivity to model parameterizations gets larger.

Figure~\ref{fig:pspectra} presents the harmonic coefficients in various centrality bins, where $v_8$ and $v_9$ are measured for the first time at the LHC energies. The magnitudes of $v_8$ and $v_9$ are compatible with $v_7$ within uncertainties. The decrease of magnitude as a function of harmonic $n$ can be characterized by a relation $v_n\propto \mathrm{e}^{-k'n^2}$ up to $v_8$. The effect might be described by viscous damping based on studies in Ref.~\cite{Staig:2011wj,Lacey:2013is}: a higher frequency waveform propagating through the medium is more strongly damped than the lower frequencies until freeze-out takes place. The relation is observed as partially broken at very high harmonics, where the damping in magnitude is no longer exponential. As predicted by the acoustic model~\cite{Staig:2011wj}, the harmonic phase oscillation itself might contribute to the magnitude of the flow coefficients at the freeze-out, which could result in enhancement of magnitude at very high harmonics. %observed as harmonic peaks at specific harmonics. One such peak is predicted at $v_9$, of which a hint is reflected by the measurements.

%Figure~\ref{fig:pspectra_hydro} shows the model calculations as a function of harmonic order in a logarithmic scale. The comparisons include calculations with constant and temperature dependent parameterizations for $\eta/s$ and $\zeta/s$, and initial condition models based on physical assumptions or Bayesian analysis. All models reproduce the hydrodynamic damping observed in Fig.~\ref{fig:pspectra}, and the slope of the calculations is dependent on the model parameterizations. Generally, a higher $\eta/s$ results more damped higher harmonic orders. Among all calculations, the rendering of the slope agrees with each other.

%\begin{figure}[tbp]
%	\centering
%	\hspace{-1.28em}\begin{overpic}[width=1.06\textwidth]{figs/vna_pspectra}
%	\put(7,1){\includegraphics[width=0.3\textwidth]{figs/vna_pspectra-pre.png}}
%	\end{overpic}
%	\caption{$v_n$ as a function of the harmonic order $n$ for various centrality intervals.}
% 	\label{fig:pspectra}
%\end{figure}

%\begin{figure}[tbp]
%	\centering
%	\includegraphics[width=1.0\textwidth]{figs/fit_pspectra_theory_2cent_nodata_nofit}
%	\caption{Hydrodynamic calculations in logarithmic scale, showing the slope of the acoustic damping up to $v_8$. For each calculation, the magnitude of the harmonics have been scaled by $10^a$ for visual clarity.}
%	\label{fig:pspectra_hydro}
%\end{figure}

\begin{figure}[tbp]
	\centering
	%\hspace{0em}\includegraphics[width=0.80\textwidth]{figs/chi_523}
	\hspace{-2.8em}\begin{overpic}[width=0.75\textwidth]{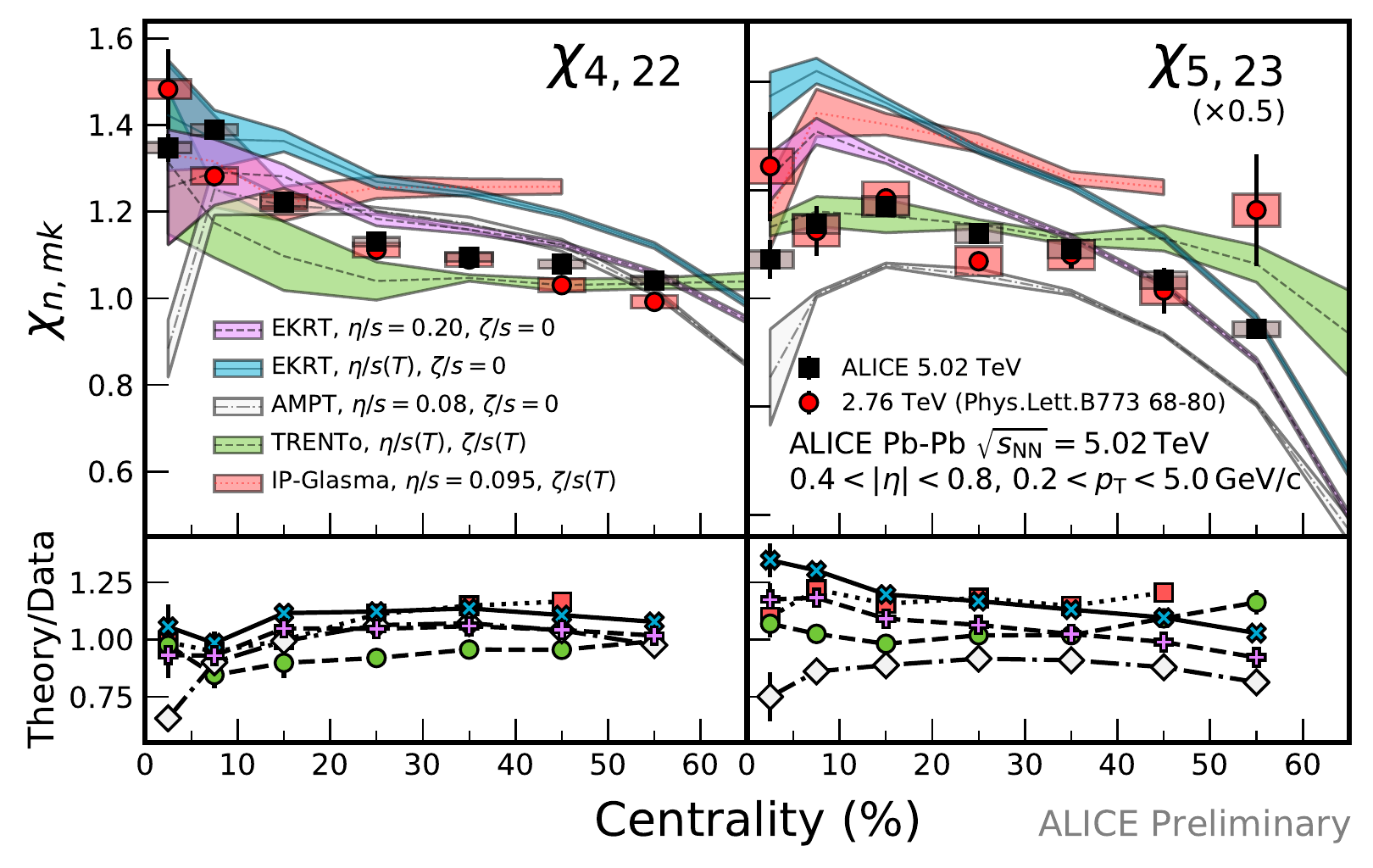}
		\put(0,1.5){\includegraphics[width=0.18\textwidth]{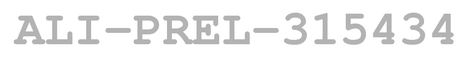}}
	\end{overpic}
	\caption{Non-linear flow mode coefficients. The new measurements are presented as black squares, while the lower energy results are in red~\cite{Acharya:2017zfg}. The model calculations are shown in various colored bands.}
 	\label{fig:chi}
\end{figure}

The results of the non-linear flow mode coefficients $\chi_{4,22}$ and $\chi_{5,23}$ are presented in Fig.~\ref{fig:chi}. For both coefficients, the overall centrality dependence is subtly decreasing, and the $\chi_{5,23}$ is about twice in magnitude compared to $\chi_{4,22}$. The results agree with those at $\sqrt{s_\mathrm{NN}}=2.76\,\mathrm{TeV}$. Discrepancies between data and model calculations are clearly visible. EKRT($\eta/s=0.2$) and \trento{} have the best agreement with the data. EKRT($\eta/s(T)$) and IP-Glasma both overestimate the data, while AMPT underestimates $\chi_{5,23}$. The sensitivity of the higher order $\chi_{5,23}$ to model parameterizations at freeze-out temperature is more prominent than for $\chi_{4,22}$, and is further pronounced in even higher harmonics~\cite{Yan:2015jma}. None of the models reproduce the centrality dependence of the non-linear flow mode coefficients within uncertainties.%perfectly capture the centrality dependence of the non-linear flow mode coefficients, closest being \trento{}.

\begin{figure}[tbp]
	\centering
	%\hspace{0em}\includegraphics[width=0.80\textwidth]{figs/chi_523}
	\includegraphics[width=0.36\textwidth]{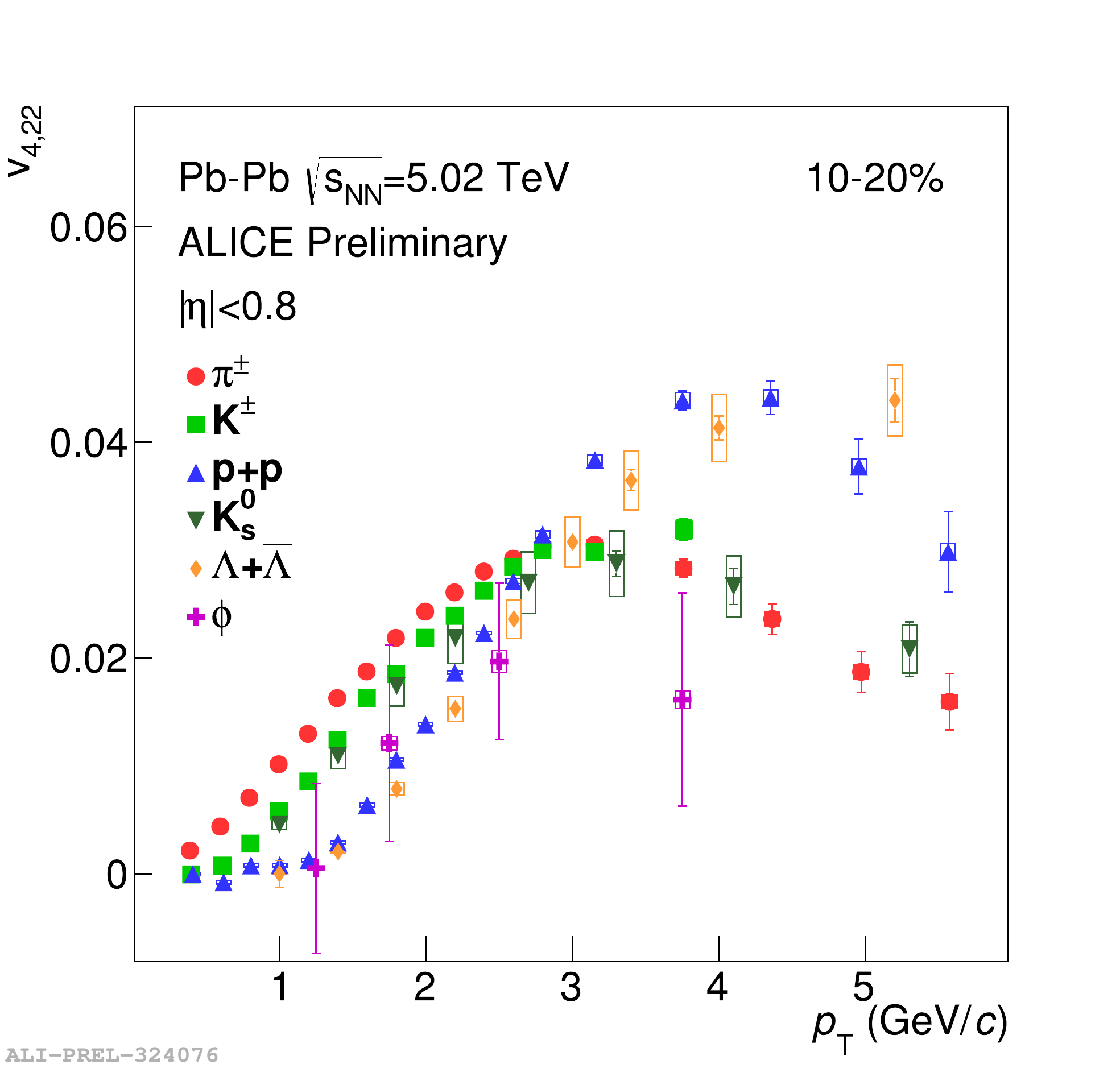}
	\includegraphics[width=0.36\textwidth]{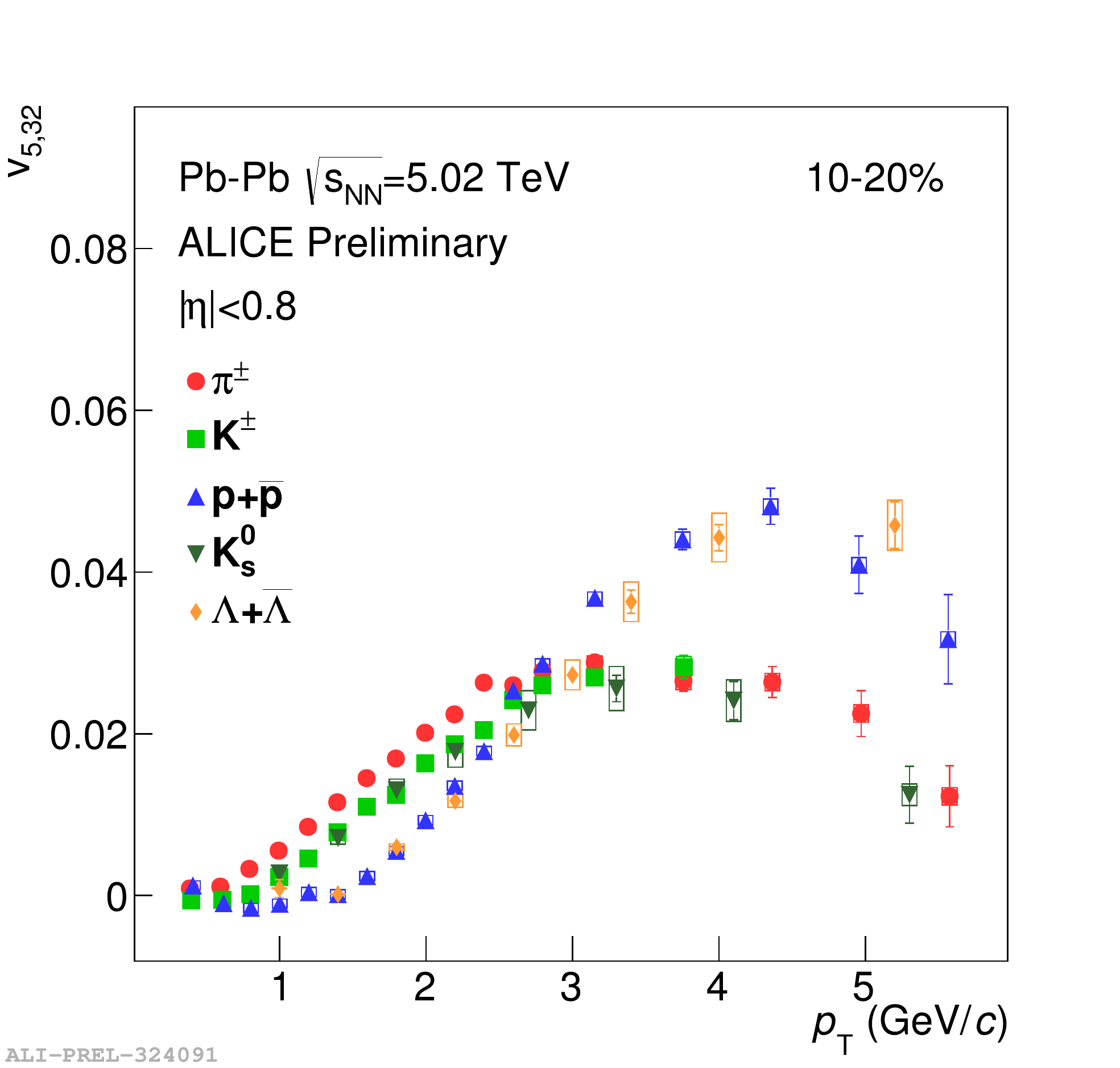}\\
	\includegraphics[width=0.36\textwidth]{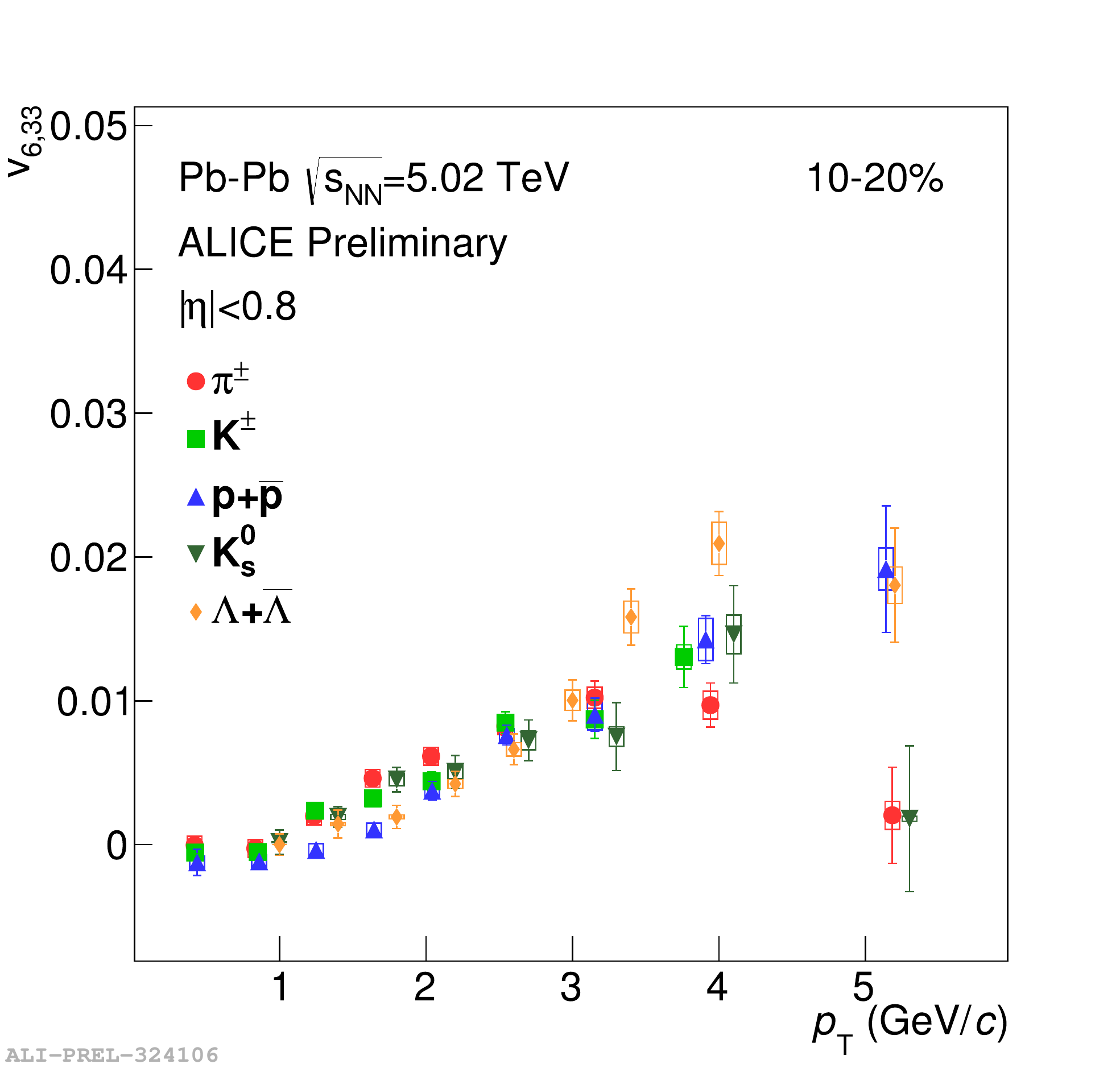}
	\includegraphics[width=0.36\textwidth]{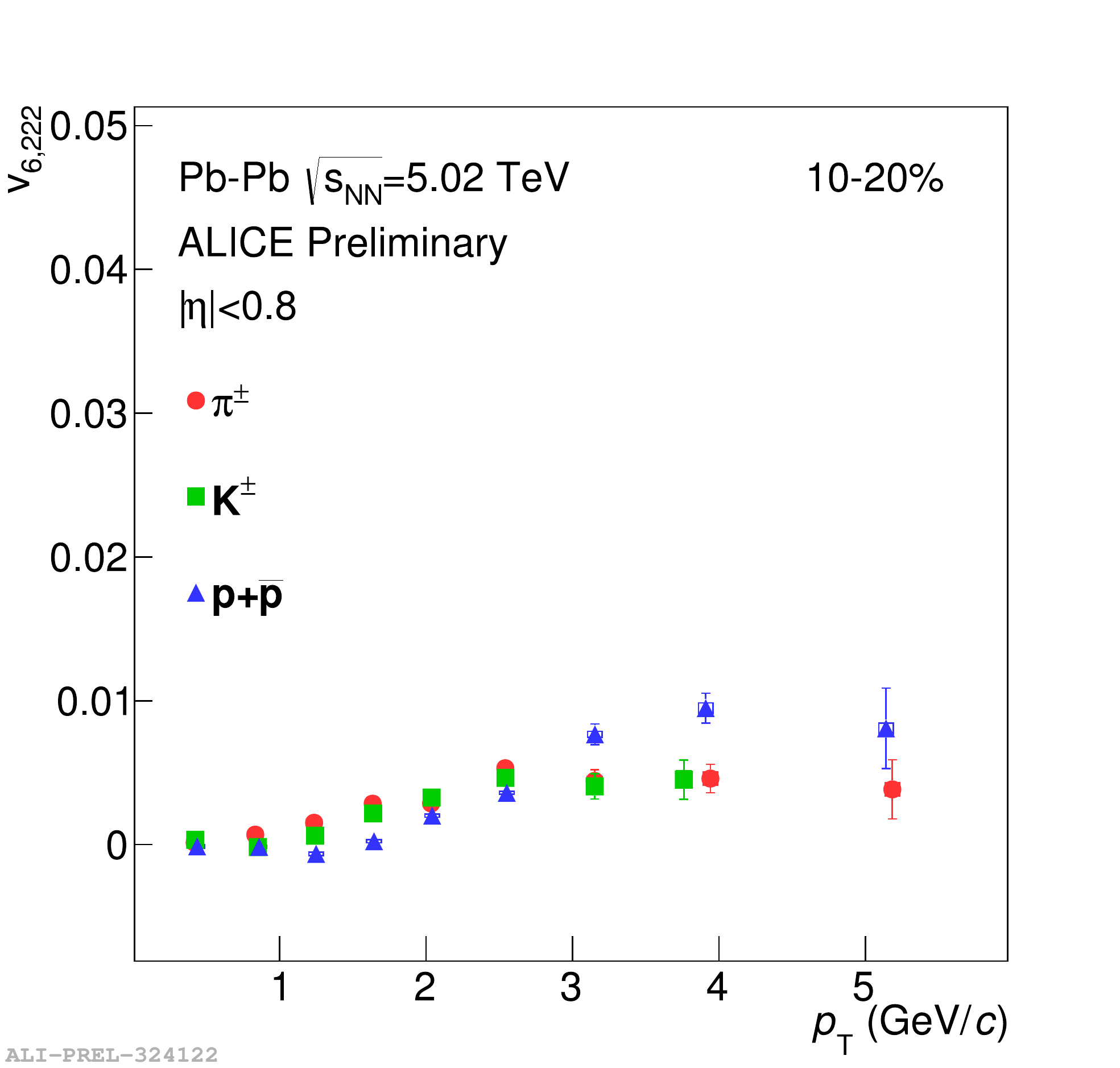}
	\caption{Non-linear flow modes of identified hadrons for $v_{4,22}$, $v_{5,23}$, $v_{6,33}$ and $v_{6,222}$.}
 	\label{fig:vnmpid}
\end{figure}

Finally, Fig.~\ref{fig:vnmpid} presents the $p_\mathrm{T}$-differential non-linear flow modes for the identified hadrons in 10-20\% centrality class. For all $v_{n,mk}$, a mass ordering $v_{n,mk}^{\pi^\pm}>v_{n,mk}^\mathrm{K}>v_{n,mk}^p\approx v_{n,mk}^\Lambda\approx v_{n,mk}^\phi$ is observed up to $p_\mathrm{T}=2.5\,\mathrm{GeV}/c$, known to be caused by an interplay between the radial flow and the non-linear response. At $p_\mathrm{T}>2.5\,\mathrm{GeV}/c$, particle type grouping is recognized with baryon flow larger than the meson flow $v_{n,mk}^\Lambda\approx v_{n,mk}^p>v_{n,mk}^{\pi^\pm}\approx v_{n,mk}^\mathrm{K}$. Such behaviour has also been observed for flow coefficients $v_n$ in \cite{Abelev:2014pua}, which implies the particle production primarily by quark coalescence, meaning that the flow is generated during the partonic phase of the collision process. %This implies the quark coalescence to be the primary mechanism in particle production.

\section{Summary}
\label{sec:summary}

\noindent The results for the higher order anisotropic flow coefficients $v_n$ and the non-linear flow mode coefficients $\chi_{n,mk}$ are presented up to ninth and fifth harmonic, respectively.
%A clear harmonic ordering is observed for the viscous damping of the flow coefficients up to $n=7$. For $n>7$, the ordering is broken with the subsequent harmonics being enhanced by a small magnitude.
The magnitude of flow coefficients decreases exponentially as the harmonic $n$ increases up to $n=7$. For $n>7$, this is no longer clearly visible.
Future measurements of $v_n$ based on the large Pb--Pb collisions data set taken in 2018 will allow to investigate damping of harmonics at $n>7$. The measurements of the non-linear flow mode coefficients show the increased non-linear response in higher harmonic. Out of the presented model comparisons EKRT($\eta/s=0$) and \trento{} reproduce the data best. However, none of the models accurately reproduces the magnitude and centrality dependence of the data, which means that further optimization should be done on the model parameters. Measurements of the $p_\mathrm{T}$-differential non-linear flow modes for identified hadrons were presented up to the sixth harmonic. The measurements show a clear mass ordering at $p_\mathrm{T} < 2.5\,\mathrm{GeV}/c$. At higher $p_\mathrm{T}$ range, particle type grouping is observed. Measurements of identified flow are an important additional constraint, as non-linear flow of various particle species can provide information especially on late-stage hadronic interactions and mechanisms of particle production. These new measurements allow the transport properties of the QGP to be better constrained.


%% References


\bibliographystyle{elsarticle-num}
\bibliography{references}

\begin{thebibliography}{10}
\expandafter\ifx\csname url\endcsname\relax
  \def\url#1{\texttt{#1}}\fi
\expandafter\ifx\csname urlprefix\endcsname\relax\def\urlprefix{URL }\fi
\expandafter\ifx\csname href\endcsname\relax
  \def\href#1#2{#2} \def\path#1{#1}\fi

\bibitem{Voloshin:2008dg}
S.~A. Voloshin, A.~M. Poskanzer, R.~Snellings, {Collective phenomena in
  non-central nuclear collisions}, Landolt-Bornstein 23 (2010) 293--333.
\newblock \href {http://arxiv.org/abs/0809.2949} {\path{arXiv:0809.2949}},
  \href {http://dx.doi.org/10.1007/978-3-642-01539-7_10}
  {\path{doi:10.1007/978-3-642-01539-7_10}}.

\bibitem{Gardim:2011xv}
F.~G. Gardim, F.~Grassi, M.~Luzum, J.-Y. Ollitrault, {Mapping the hydrodynamic
  response to the initial geometry in heavy-ion collisions}, Phys. Rev. C85
  (2012) 024908.
\newblock \href {http://arxiv.org/abs/1111.6538} {\path{arXiv:1111.6538}},
  \href {http://dx.doi.org/10.1103/PhysRevC.85.024908}
  {\path{doi:10.1103/PhysRevC.85.024908}}.

\bibitem{Acharya:2020taj}
S.~Acharya, et~al., {Linear and non-linear flow modes of charged hadrons in
  Pb-Pb collisions at $\sqrt{s_{\rm{NN}}}$ = 5.02 TeV}\href
  {http://arxiv.org/abs/2002.00633} {\path{arXiv:2002.00633}}.

\bibitem{Acharya:2019uia}
S.~Acharya, et~al., {Non-linear flow modes of identified particles in Pb-Pb
  collisions at $\sqrt{s_{\rm{NN}}}$ = 5.02 TeV}\href
  {http://arxiv.org/abs/1912.00740} {\path{arXiv:1912.00740}}.

\bibitem{Aamodt:2008zz}
K.~Aamodt, et~al., {The ALICE experiment at the CERN LHC}, JINST 3 (2008)
  S08002.
\newblock \href {http://dx.doi.org/10.1088/1748-0221/3/08/S08002}
  {\path{doi:10.1088/1748-0221/3/08/S08002}}.

\bibitem{Bilandzic:2013kga}
A.~Bilandzic, C.~H. Christensen, K.~Gulbrandsen, A.~Hansen, Y.~Zhou, {Generic
  framework for anisotropic flow analyses with multiparticle azimuthal
  correlations}, Phys. Rev. C89~(6) (2014) 064904.
\newblock \href {http://arxiv.org/abs/1312.3572} {\path{arXiv:1312.3572}},
  \href {http://dx.doi.org/10.1103/PhysRevC.89.064904}
  {\path{doi:10.1103/PhysRevC.89.064904}}.

\bibitem{Staig:2011wj}
P.~Staig, E.~Shuryak, {The Fate of the Initial State Fluctuations in Heavy Ion
  Collisions. III The Second Act of Hydrodynamics}, Phys. Rev. C84 (2011)
  044912.
\newblock \href {http://arxiv.org/abs/1105.0676} {\path{arXiv:1105.0676}},
  \href {http://dx.doi.org/10.1103/PhysRevC.84.044912}
  {\path{doi:10.1103/PhysRevC.84.044912}}.

\bibitem{Lacey:2013is}
R.~A. Lacey, Y.~Gu, X.~Gong, D.~Reynolds, N.~N. Ajitanand, J.~M. Alexander,
  A.~Mwai, A.~Taranenko, {Is anisotropic flow really acoustic?}\href
  {http://arxiv.org/abs/1301.0165} {\path{arXiv:1301.0165}}.

\bibitem{Yan:2015jma}
L.~Yan, J.-Y. Ollitrault, {$\nu_4, \nu_5, \nu_6, \nu_7$: nonlinear hydrodynamic
  response versus LHC data}, Phys. Lett. B744 (2015) 82--87.
\newblock \href {http://arxiv.org/abs/1502.02502} {\path{arXiv:1502.02502}},
  \href {http://dx.doi.org/10.1016/j.physletb.2015.03.040}
  {\path{doi:10.1016/j.physletb.2015.03.040}}.

\bibitem{Abelev:2014pua}
B.~B. Abelev, et~al., {Elliptic flow of identified hadrons in Pb--Pb collisions
  at $ \sqrt{s_{\mathrm{NN}}}=2.76 $ TeV}, JHEP 06 (2015) 190.
\newblock \href {http://arxiv.org/abs/1405.4632} {\path{arXiv:1405.4632}},
  \href {http://dx.doi.org/10.1007/JHEP06(2015)190}
  {\path{doi:10.1007/JHEP06(2015)190}}.

\bibitem{Acharya:2017zfg}
S.~Acharya, et~al., {Linear and non-linear flow modes in Pb-Pb collisions at
  $\sqrt{s_{\rm NN}} =$ 2.76 TeV}, Phys. Lett. B773 (2017) 68--80.
\newblock \href {http://arxiv.org/abs/1705.04377} {\path{arXiv:1705.04377}},
  \href {http://dx.doi.org/10.1016/j.physletb.2017.07.060}
  {\path{doi:10.1016/j.physletb.2017.07.060}}.

\end{thebibliography}






% \end{thebibliography}

\end{document}